\documentclass[twoside,twocolumn,9pt]{article}
\usepackage{extsizes}
\usepackage[super,sort&compress,comma]{natbib} 
\usepackage[version=3]{mhchem}
\usepackage[left=1.5cm, right=1.5cm, top=1.785cm, bottom=2.0cm]{geometry}
\usepackage{balance}
\usepackage{widetext}
\usepackage{times,mathptmx}
\usepackage{sectsty}
\usepackage{graphicx} 
\usepackage{lastpage}
\usepackage[format=plain,justification=raggedright,singlelinecheck=false,font={stretch=1.125,small,sf},labelfont=bf,labelsep=space]{caption}
\usepackage{float}
\usepackage{fancyhdr}
\usepackage{fnpos}
\usepackage[english]{babel}
\usepackage{array}
\usepackage{droidsans}
\usepackage{charter}
\usepackage[T1]{fontenc}
\usepackage[usenames,dvipsnames]{xcolor}
\usepackage{setspace}
\usepackage[compact]{titlesec}

\usepackage{epstopdf}

\definecolor{cream}{RGB}{222,217,201}

\begin{document}

\pagestyle{fancy}
\thispagestyle{plain}
\fancypagestyle{plain}{

\fancyhead[C]{\includegraphics[width=18.5cm]{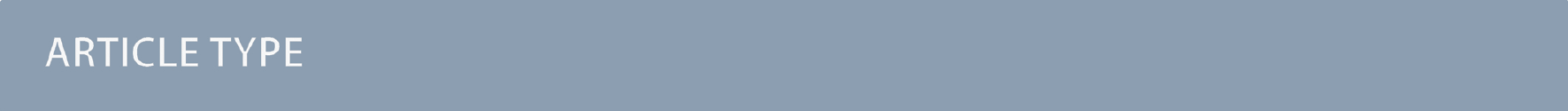}}
\fancyhead[L]{\hspace{0cm}\vspace{1.5cm}\includegraphics[height=30pt]{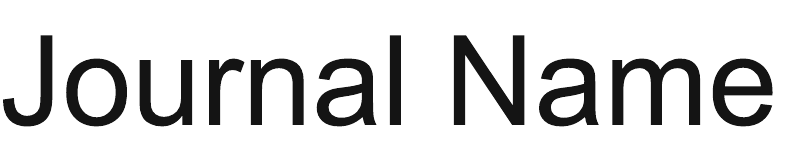}}
\fancyhead[R]{\hspace{0cm}\vspace{1.7cm}\includegraphics[height=55pt]{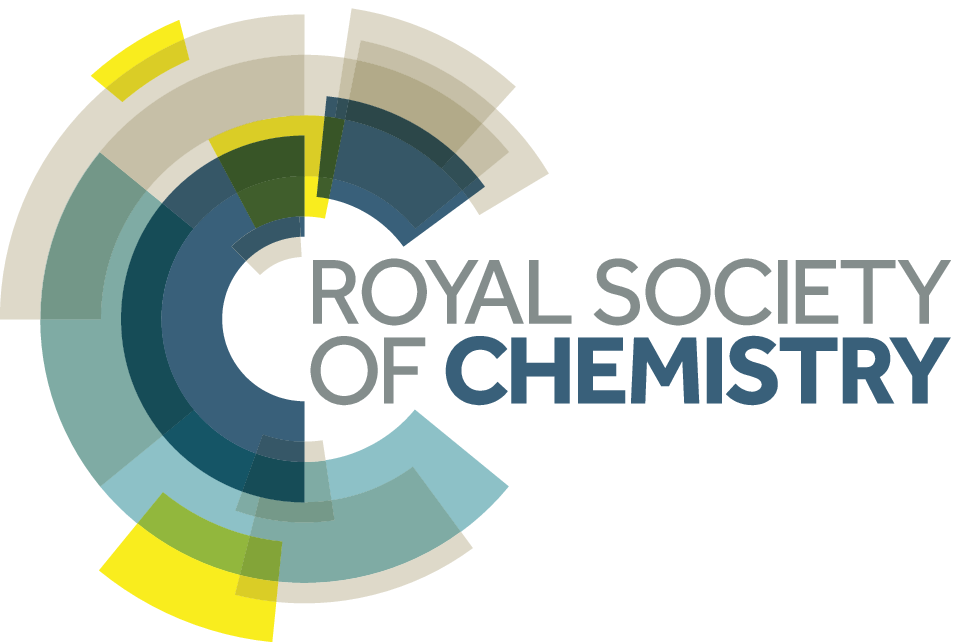}}
\renewcommand{\headrulewidth}{0pt}
}

\makeFNbottom
\makeatletter
\renewcommand\LARGE{\@setfontsize\LARGE{15pt}{17}}
\renewcommand\Large{\@setfontsize\Large{12pt}{14}}
\renewcommand\large{\@setfontsize\large{10pt}{12}}
\renewcommand\footnotesize{\@setfontsize\footnotesize{7pt}{10}}

\makeatother

\renewcommand{\thefootnote}{\fnsymbol{footnote}}
\renewcommand\footnoterule{\vspace*{1pt}%
\color{cream}\hrule width 3.5in height 0.4pt \color{black}\vspace*{5pt}} 
\setcounter{secnumdepth}{5}

\makeatletter 
\renewcommand\@biblabel[1]{#1}            
\renewcommand\@makefntext[1]%
{\noindent\makebox[0pt][r]{\@thefnmark\,}#1}
\makeatother 
\renewcommand{\figurename}{\small{Fig.}~}
\sectionfont{\sffamily\Large}
\subsectionfont{\normalsize}
\subsubsectionfont{\bf}
\setstretch{1.125} 
\setlength{\skip\footins}{0.8cm}
\setlength{\footnotesep}{0.25cm}
\setlength{\jot}{10pt}
\titlespacing*{\section}{0pt}{4pt}{4pt}
\titlespacing*{\subsection}{0pt}{15pt}{1pt}

\fancyfoot{}
\fancyfoot[LO,RE]{\vspace{-7.1pt}\includegraphics[height=9pt]{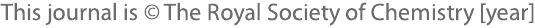}}
\fancyfoot[CO]{\vspace{-7.1pt}\hspace{13.2cm}\includegraphics{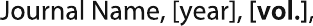}}
\fancyfoot[CE]{\vspace{-7.2pt}\hspace{-14.2cm}\includegraphics{head_foot/RF}}
\fancyfoot[RO]{\footnotesize{\sffamily{1--\pageref{LastPage} ~\textbar  \hspace{2pt}\thepage}}}
\fancyfoot[LE]{\footnotesize{\sffamily{\thepage~\textbar\hspace{3.45cm} 1--\pageref{LastPage}}}}
\fancyhead{}
\renewcommand{\headrulewidth}{0pt} 
\renewcommand{\footrulewidth}{0pt}
\setlength{\arrayrulewidth}{1pt}
\setlength{\columnsep}{6.5mm}
\setlength\bibsep{1pt}

\makeatletter 
\newlength{\figrulesep} 
\setlength{\figrulesep}{0.5\textfloatsep} 

\newcommand{\topfigrule}{\vspace*{-1pt}%
\noindent{\color{cream}\rule[-\figrulesep]{\columnwidth}{1.5pt}} }

\newcommand{\botfigrule}{\vspace*{-2pt}%
\noindent{\color{cream}\rule[\figrulesep]{\columnwidth}{1.5pt}} }

\newcommand{\dblfigrule}{\vspace*{-1pt}%
\noindent{\color{cream}\rule[-\figrulesep]{\textwidth}{1.5pt}} }

\makeatother

\twocolumn[
  \begin{@twocolumnfalse}
\vspace{3cm}
\sffamily
\begin{tabular}{m{4.5cm} p{13.5cm} }

\includegraphics{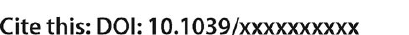} & \noindent\LARGE{\textbf{Theory of 2D crystals: graphene and beyond}} \\
\vspace{0.3cm} & \vspace{0.3cm} \\

 & \noindent\large{Rafael Rold\'an,$^{\ast}$\textit{$^{a}$} Luca Chirolli,\textit{$^{b}$} Elsa Prada,\textit{$^{c}$} Jose Angel Silva-Guill\'en,\textit{$^{b}$} Pablo San-Jose\textit{$^{a}$} and Francisco Guinea\textit{$^{b,d}$}} \\

\includegraphics{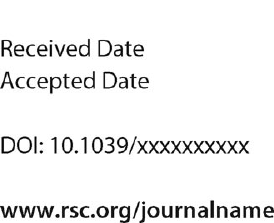} & \noindent\normalsize{ This tutorial review presents an overview of the basic theoretical aspects of two-dimensional (2D) crystals. We revise essential aspects of graphene and the new families of semiconducting 2D materials, like transition metal dichalcogenides or black phosphorus. Minimal theoretical models for various materials are presented. Some of the exciting new possibilities offered by 2D crystals are discussed, such as manipulation and control of quantum degrees of freedom (spin and pseudospin), confinement of excitons, control of the electronic and optical properties with strain engineering, or unconventional superconducting phases.}\\

&\\

& {\bf Key learning points:} Novel two dimensional crystals. Models for the electronic structure. Spin and pseudospin. Excitons and optical absorption. Superconductivity.

\end{tabular}

\end{@twocolumnfalse} \vspace{0.6cm}

  ]

\renewcommand*\rmdefault{bch}\normalfont\upshape
\rmfamily
\section*{}
\vspace{-1cm}


\footnotetext{\textit{$^{a}$~Instituto de Ciencia de Materiales de Madrid, ICMM-CSIC, Cantoblanco, E-28049 Madrid, Spain. Fax: +34 913720623; Tel: +34 913349000; E-mail: rroldan@icmm.csic.es}}
\footnotetext{\textit{$^{b}$~Fundaci\'on IMDEA Nanociencia, C/Faraday 9, Campus Cantoblanco, 28049 Madrid, Spain}}
\footnotetext{\textit{$^{c}$~Departamento de F\'isica de la Materia Condensada, Condensed Matter Physics Center (IFIMAC) 
 \& Instituto Nicolas Cabrera, Universidad Aut\'onoma de Madrid, E-28049 Madrid, Spain}}
\footnotetext{\textit{$^{d}$~Department of Physics and Astronomy, University of Manchester, Oxford Road, Manchester M13 9PL, UK}}

\section{Introduction}

Graphene is the first truly 2D crystal that has been isolated in a controlled manner, initiating a field of research known as "2D materials".\cite{Novoselov_PNAS_2005} 
2D crystals are materials of atomic thickness that, as a result of their reduced dimensionality, exhibit unique physical and chemical properties that strongly differ from their 3D counterparts. If the crystalline structure of a 3D crystal is preserved as its thickness is reduced down to atomic scales, as happens with layered materials, it typically exhibits dramatic changes in its physical properties.  For example, only when graphene is isolated from 3D graphite into its one-atom-thick form, do its carriers behave as massless relativistic electrons. Another example is MoS$_2$, which in its monolayer form presents a direct band gap and spin-polarized valleys, two key features that make this 2D semiconductor much better suited for photonics and optoelectronics than its 3D version. From a fundamental point of view, moreover, 2D is a critical dimensionality for many physical effects, and marks a threshold wherein thermal and quantum fluctuations acquire a much more dominant role than in 3D. This alone often makes the physics of 2D crystals highly non-trivial.

In more practical terms, the all-surface nature of 2D crystals exposes them much more directly to influences of all sorts from the environment. As a consequence, their electronic and optical properties can be tuned with a particularly high degree of flexibility.  Parameters associated to the electronic structure of crystals, like effective masses, Fermi energy, Fermi velocity or band gap, can be efficiently tuned by controlling the number of layers, by chemical functionalization, by gating or by applying strain to the samples or the substrate on which 2D materials are deposited. As compared to other 2D systems such as conventional thin films, 2D crystals also exhibit a much higher quality and overall coherence, as their strong covalent in-plane bonds allow them to keep disorder under control as their thickness is reduced.

Graphene is the best studied among 2D materials, with recognised hallmark properties that make it particularly attractive, both from a fundamental viewpoint and because of its potential applications.\cite{Castro-Neto_RMP_2009} However, it lacks a band gap, which is necessary to switch between insulating and metallic states, making graphene unsuitable for some electronic devices. Furthermore, a band gap in the visible or infrared range of the spectrum is also required for solar cell and telecommunication applications. Consequently, significant efforts have been devoted to identifying possible 2D semiconducting crystals.\cite{Castellanos_NP_2016} Several classes of layered compounds have received attention recently, including hexagonal boron nitride (h-BN), silicene, MoS$_2$, black phosphorus (BP), etc. Single layers of h-BN are stable insulators with a large gap. Conversely, silicene is highly unstable because single layers can react with air. Among the stable 2D crystals with semiconducting behaviour, some of the best known are transition metal dichalcogenides (TMDs),\cite{Liu_CSR_2015}  which also exhibit strong spin-orbit coupling (SOC) effects. BP is another layered crystal that has been recently synthesized in its single layer form, also known as phosphorene. BP is a stable allotrope of phosphorus and an elemental semiconductor, with a high degree of anisotropy in its electronic and optical properties.\cite{Castellanos_JPCL_2015} Much less understood are other families of transition metal trichalcogenides like TiS$_3$, single layer Sb (antimonene) or monochalcogenides like GeSe, that have been recently synthesised in their single layer form, joining the growing catalogue of 2D materials. Herein, we review current knowledge on the physical properties of graphene and related 2D crystals. We survey their main electronic and structural features and provide  minimal theoretical models that capture their low energy physics. Finally, we discuss some of the novel possibilities afforded by 2D crystals, including spintronics and valleytronics, control of excitons, strain engineering and new aspects of superconductivity.  

\section{General description}
\label{Sec:description}

In this section we review the crystalline order and electronic band structure of the most relevant 2D materials, and highlight some of their main features.

\begin{figure*}[h!]
\centering
  \includegraphics[width=\textwidth]{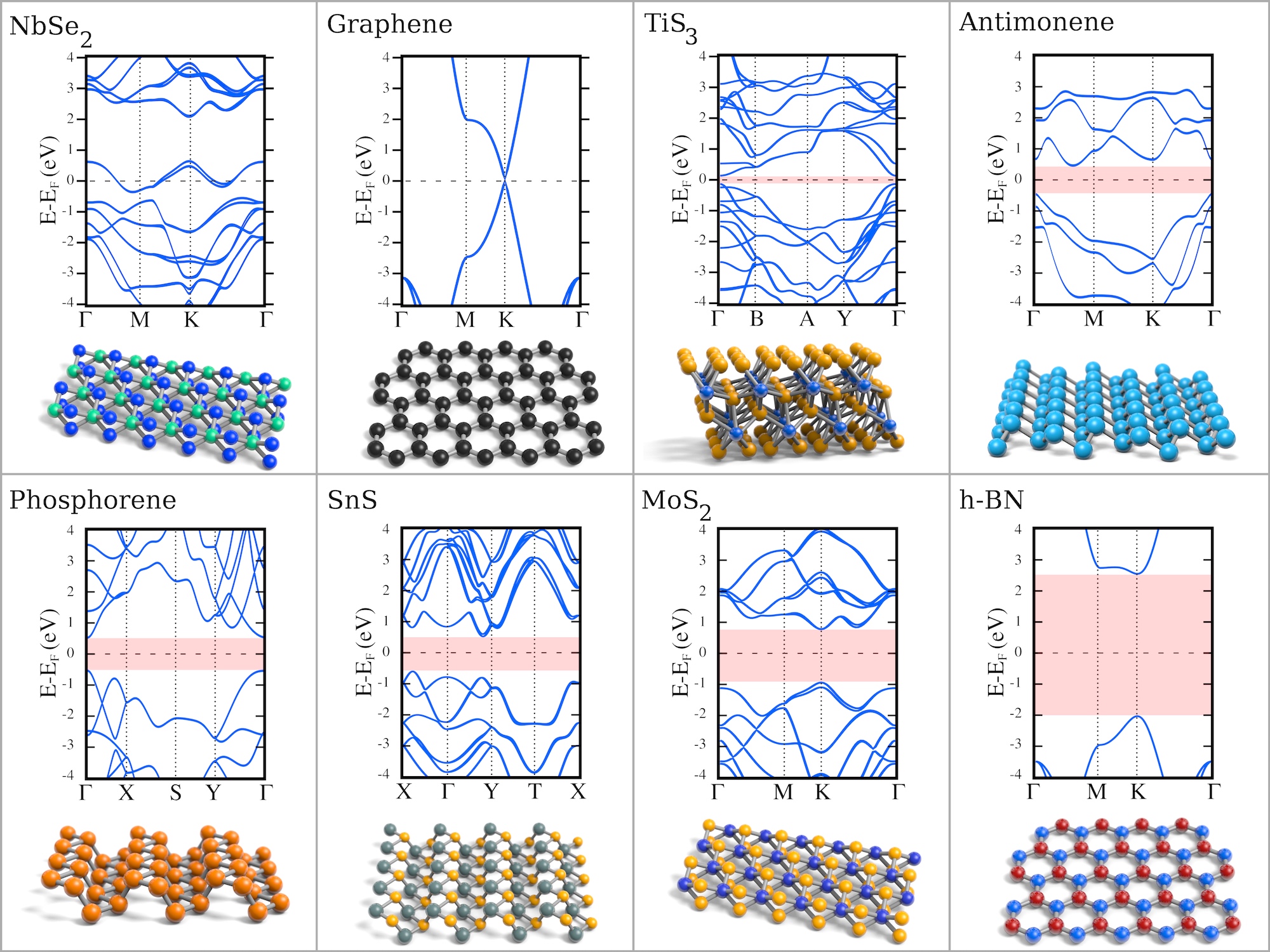}
  \caption{Crystalline lattice and band structures of different 2D materials, ordered by increasing size of their gap (red shadowed area) as computed within DFT without correlation effects: metallic (NbSe$_2$), semi-metallic (graphene), semiconducting (TiS$_3$, antimonene, phosphorene, SnS, MoS$_2$) and insulating (h-BN).}
  \label{Fig:Bands}
\end{figure*}

\subsection{Graphene}

Graphene is a one-atom-thick layer of carbon atoms, which when stacked constitute 3D graphite. As in other carbon structures like 0D fullerenes (C$_{60}$) or 1D carbon nanotubes, 2D graphene is a crystal composed of $sp^2$-hybridized carbon atoms arranged in a honeycomb lattice. However, the physical and chemical properties of graphene strongly differ from the other $sp^2$ and $sp^3$ carbon compounds of different dimensionality.  Its electronic bands define it as a semimetal, at the boundary between the two most frequent phases of materials, metals and insulators.  The electronic band structure of graphene is characterised by conical valence and conduction bands that touch at only two inequivalent points in the Brillouin zone (BZ), dubbed Dirac points. The low energy properties are governed by out-of-plane $p_z$-orbitals. Some of the main characteristics of graphene are:\cite{Katsnelson_Book_2012} 

\begin{itemize}

\item {\bf Massless carriers:} The characteristic linear dispersion relation of electrons in graphene in the vicinity of the Dirac points makes them behave as relativistic quasiparticles with zero effective mass (see Sec. \ref{Sec:Models}). As a result, the electronic and optical properties of graphene are completely different from those of a standard 2D electron gas with a {\it massive} parabolic dispersion relation, as e.g. Si and GaAlAs heterostructures. When graphene is exposed to a strong magnetic field, the massless character of the carriers manifests in an unconventional quantum Hall effect.\cite{Goerbig_RMP_2011} Graphene electrons can, moreover, propagate over large (micrometers) distances without scattering due to the chirality of an internal degree of freedom of carriers known as pseudospin.

\item {\bf High stiffness and impermeability:} Graphene is simultaneously flexible and extraordinarily rigid, with the highest elastic constants ever measured in any material. It can be stretched elastically up to $\sim 20\%$ without rupture. Graphene (as other 2D membranes) presents a negative thermal expansion coefficient: it shrinks with increasing temperature, due to elastic properties that are dominated by out-of-plane flexural phonons.\cite{Amorim_PR_2016} Despite its one-atom-thickness, graphene is highly impermeable to gases. 

\end{itemize}

Some of these attributes are shared by other 2D crystals that have been exfoliated after graphene.

\subsection{Transition Metal Dichalcogenides $MX_2$}

TMDs form a very large and rich family of crystals, whose members present several kinds of lattice structures and many different physical properties. The most studied are the semiconducting compounds with the general formula $MX_2$ ($M=$ Mo, W; $X=$ S, Se, Te) that are composed, in its bulk configuration, of two-dimensional $X-M-X$ layers stacked on top of each other, coupled by weak van der Waals forces. A $MX_2$ single layer is a sandwich structure in which the $M$ atoms are ordered in a triangular lattice, each of them bonded covalently to six $X$ atoms, three of them in the top and three in the bottom layer, see Fig. \ref{Fig:Bands}. Some of the main features of TMDs include:\cite{Roldan_AP_2014}

\begin{itemize}

\item {\bf Thickness dependence of the nature of the gap:} TMDs have an electronic band structure which strongly depends on the number of layers. Single layer TMDs are direct gap semiconductors, with a gap ($\sim 1.9$~eV) located at the K and K' points of the hexagonal BZ. The energy of the band gap lies in the visible range of the spectrum. Multilayer samples are indirect gap semiconductors (of $\sim 1.3$~eV gap) with the maximum of the valence band at $\Gamma$ and the minimum of the conduction band at an intermediate point between $\Gamma$ and K. 

\item {\bf Multi-Orbital character:} TMDs have valence and conduction bands with a rich and complex orbital character. The edge of the valence band is formed by a mixture of $d_{x^2-y^2}$ and $d_{xy}$ orbitals of the metal $M$, and $p_x$ and $p_y$ orbitals of the chalcogen atom $X$.  The edge of the conduction band, on the other hand, is formed by a combination of $d_{3z^2-r^2}$ of $M$, plus some minor contribution of chalcogen $p_x$ and $p_y$ orbitals.  

\item {\bf Strong spin-orbit coupling:} TMDs present a strong SOC which, together with the lack of inversion symmetry, leads to a splitting of the valence band of $\sim 140$ meV (for Mo compounds) and $\sim 400$ meV (for W compounds). The conduction band is also split by a few tens of meV.

\end{itemize}

Other characteristics of TMDs that we will discuss in Sec. \ref{Sec:Novel} include the coexistence of spin and valley Hall effects, the possibility of tuning the band gap by different means, like strain engineering or application of external electric fields, generation of highly stable excitons (electron-hole pairs bonded by Coulomb interaction), or the emergence of superconductivity in highly doped samples.   

While the most commonly studied TMDs in the single-layer form are those with TMs from group VI-B and are semiconducting, those with TMs from groups IV-B and V-B have also been studied. The ones from group V-B with trigonal prismatic coordination are metallic.\cite{Chhowalla_NC_2013} The change from semiconductor to metallic phase can be easily explained, since group V-B TMs have one valence electron less than the TMs from group VI-B and, therefore, do not have enough electrons in the unit cell to occupy the valence band like the group VI-B TMDs. The most studied TMDs from group V-B are NbS$_2$, NbSe$_2$ and TaS$_2$. These materials present electronic correlation phenomena such as superconductivity and even, in the case of NbSe$_2$ and TaS$_2$, a charge density wave phase that can compete with the superconducting state. Those of group IV-B are semimetals and, for example, doped TiSe$_2$ exhibits a charge density wave as well as superconductivity.

\subsection{Phosphorene}

Black phosphorus (BP) is another layered material that has been recently synthesized in its single layer form, also known as phosphorene.\cite{Castellanos_JPCL_2015} BP is a stable allotrope of phosphorus and an elemental semiconductor. As in graphene, each P atom in single layer phosphorene is coordinated to three neighbouring atoms. But contrary to graphene, the orbital hybridization is $sp^3$ like, yielding a crystal structure characterized by a puckered honeycomb lattice with a high level of structural anisotropy (Fig. \ref{Fig:Bands}). The edges of the valence and conduction bands are formed from $p_z$ orbitals.\cite{Rudenko_PRB_2015} Some characteristics of BP to be highlighted are:

\begin{itemize}

\item {\bf Evolution of the band gap with thickness and applied strain:} The gap in single-layer as well as in multi-layer samples is direct and located at the $\Gamma$ point of the BZ. However, while the energy of the gap in bulk BP is $\sim 0.3$~eV, its value increases with decreasing the layer number to $\sim 1.5$~eV for a single layer phosphorene. For a given sample thickness, the band gap is extremely sensitive to external strain.\cite{Quereda_NL_2016} Therefore, BP provides a high feasibility for photonics and optoelectronics devices that can operate at different frequencies. BP is, in this sense, complementary to some of the most studied 2D crystals, namely graphene and TMDs, that have band gaps ranging from zero in graphene and the $\sim 2$~eV in TMDs ($\sim 2.5$~eV if we include correlation effects).

\item{\bf In-plane anisotropy:} The peculiar puckered structure of BP layers leads to highly anisotropic optical and electronic properties in-plane. The non-isotropic band dispersion also yields highly anisotropic excitons and plasmons.\cite{Low_NM_2017}

\item{\bf Semiconductor-to-semimetal transition:} Thanks to the strong response of BP to electric and strain fields, it is possible to drive a semiconductor-to-semimetal transition in this material with the appearance of a pair of Dirac-like cones in the spectrum, similarly to graphene.\cite{Yuan_PRB_2016,Dutreix_PRB_2016} Such transition is accompanied by a change in the topology of the system, due to generation of $\pm \pi$ Berry phases around the Dirac points. 

\end{itemize}

\subsection{Hexagonal boron nitride h-BN}
h-BN is a band insulator with B and N atoms sitting on different sublattices of a honeycomb lattice. It presents a direct band gap of $\sim 4.5$~eV located at the K and K' points of the BZ. Few layer h-BN is widely used in 2D materials research because it is an excellent substrate to support and encapsulate graphene, black phosphorus, etc. Due to its atomic flatness and very low concentration of trapped charges, the use of h-BN as dielectric substrate instead of, e.g. SiO$_2$, considerably enhances graphene mobility with the corresponding improvement of device performance.  On the other hand, due to its small lattice mismatch with graphene ($\sim 1.8\%$), graphene on h-BN is a perfect platform to study moir\'e pattern effects on the electronic and optical properties.\cite{Yankowitz_NC_2016} h-BN is also a fundamental piece in so called van der Waals heterostructures, which is the name commonly used to refer to heterostructures and devices whose properties are engineered by stacking different 2D crystals, with the desired properties, on top of each other.\cite{Geim_N_2013}  Furthermore, h-BN is interesting in its own right as it supports mixed photon-phonon modes (polaritons) with an hyperbolic dispersion in a finite range of frequencies. These modes lead to highly confined electric fields.\cite{Caldwell_NC_2014}

\subsection{Monochalcogenides $MX$}

Group IV monochalcogenides with the common formula $MX$ ($M=$ Ge, Sn and $X=$ S, Se) are bi-elemental crystals that, as phosphorene, presents a puckered orthorhombic structure (Fig. \ref{Fig:Bands}). They are indirect gap semiconductors with a quasiparticle gap (including correlation effects)  ranging from  $1.2$~eV to $2.7$~eV.  As TMDs. Unlike phosphorene, they are not inversion-symmetric crystals. Therefore the monochalcogenides present piezoelectric properties (i.e. possibility to convert mechanical to electric energy), with piezoelectric coefficients that can be two orders of magnitude larger than in MoS$_2$ or h-BN. Indeed, their puckered structure makes them very soft along the armchair direction, a property that can further improve the piezoelectric performance.\cite{Fei_APL_2015}

\subsection{Antimonene}

Single layer antimonene is an elemental 2D crystal formed by Sb atoms ordered in a buckled honeycomb structure.\cite{Ares_AM_2016} It is an indirect gap semiconductor with strong SOC ($\lambda\sim 0.34$~eV) and a gap of $\sim 1$~eV. The valence band edge (composed of $p_x$ and $p_y$ orbitals) is located at the $\Gamma$-point of the BZ.\cite{Rudenko_PRB_2017} The bottom of the conduction band (formed by a combination of the three $p$ orbitals) is placed at a non-high symmetry point of the BZ, between $\Gamma$ and M. While single layer antimonene is a topologically trivial semiconductor, the possibility has been discussed that a transition to a topological insulator may take place upon increasing the number of layers, including the development of quantum spin Hall phases.\cite{Zhang_PRB_2012}

\subsection{Transition metal trichalcogenides $MX_3$}

Transition metal trichalcogenides (TMT) with the common formula  $MX_3$ ($M=$ Ti, Zr, Hf and $X=$ S, Se, Te) are formed from trigonal prismatic $MX_3$ chains such that two rectangular faces of a $MX_6$ trigonal prism are capped by $X$ atoms of the neighboring chains. Thus, each transition metal atom $M$ is coordinated to eight chalcogen atoms, $X$ (see Fig. \ref{Fig:Bands}). Among the TMTs family, TiS$_3$ is attracting special attention because of its direct band gap.\cite{Island_AOM_2014} Contrary to TMDs, the direct nature of the TiS$_3$ band gap is robust, with only a weak dependence of its magnitude with the number of layers or their stacking order. The reduced in-plane structural symmetry of the crystal confers a strong anisotropy to its band dispersion. Interestingly, the in-plane anisotropy is opposite for the valence and the conduction bands.\cite{Silva_2DM_2017} The ultimate origin of such different anisotropy resides in the different orbital character of the valence and conduction band edges. The top of the valence band is mostly made of the 3$p_x$ orbitals of S belonging to the `inner' sulfur atoms, dispersing along the $a$-crystallographic direction. On the other hand, the conduction band is mainly composed of $d_{3z^2-r^2}$ orbitals of Ti atoms, oriented along the $b$-direction.  The electronic and optical properties are therefore highly anisotropic, with a strong directional dependence of conductance and a large linear dichroism.  

\section{Theoretical models}\label{Sec:Models}

As with any new class of materials, theoretical models have proved to be a powerful tool to understand the properties of many 2D crystals. One may even say that the predictive role of theory is particularly important in these systems, given their rather unique combination of interrelated physical properties, of which there is hardly any precedent in more conventional materials. For example, many 2D crystals are so stable despite their reduced dimensionality that they may be suspended to behave as a robust elastic membrane of ultimate thinness, with a host of associated and highly non-trivial anharmonic effects. At the same time, a suspended 2D crystal is often of such a pristine electronic quality as to rival many semiconducting heterostructures. Crucially, moreover, the electronic and elastic degrees of freedom of the crystal are often strongly coupled. The interplay of these two physical sectors gives rise to unexpected possibilities, both fundamental and applied. Other examples of this synergetic character include the interplay between strain and optical activity in MoS${}_2$, between electronic structure, deformations and interlayer registry in various multilayers,\cite{Yankowitz_NC_2016} or between topological order, electric fields and pressure in black phosphorus.\cite{Kim:S15} Theoretical models allow to navigate all these numerous possibilities much more easily.

\subsection{A threefold approach to modelling 2D crystals}

When a clean experimental characterisation is lacking, a powerful strategy to modelling the electronic and elastic properties of 2D crystals is to combine three types of theoretical approaches, with a decreasing level of computational complexity. At the top of the stack are \emph{ab-initio} methods, that range from Density Functional Theory (DFT) to more elaborate extensions such as GW or the Bethe-Salpeter equation to capture some of the electronic correlations. These methods are very successful in some of these systems, and remain computationally viable given the relatively small number of orbitals in the unit cell of typical 2D crystals. They can often predict the single-particle bandstructure and elastic coefficients with good accuracy. 

\begin{figure}
   \centering
   \includegraphics[width=1\columnwidth]{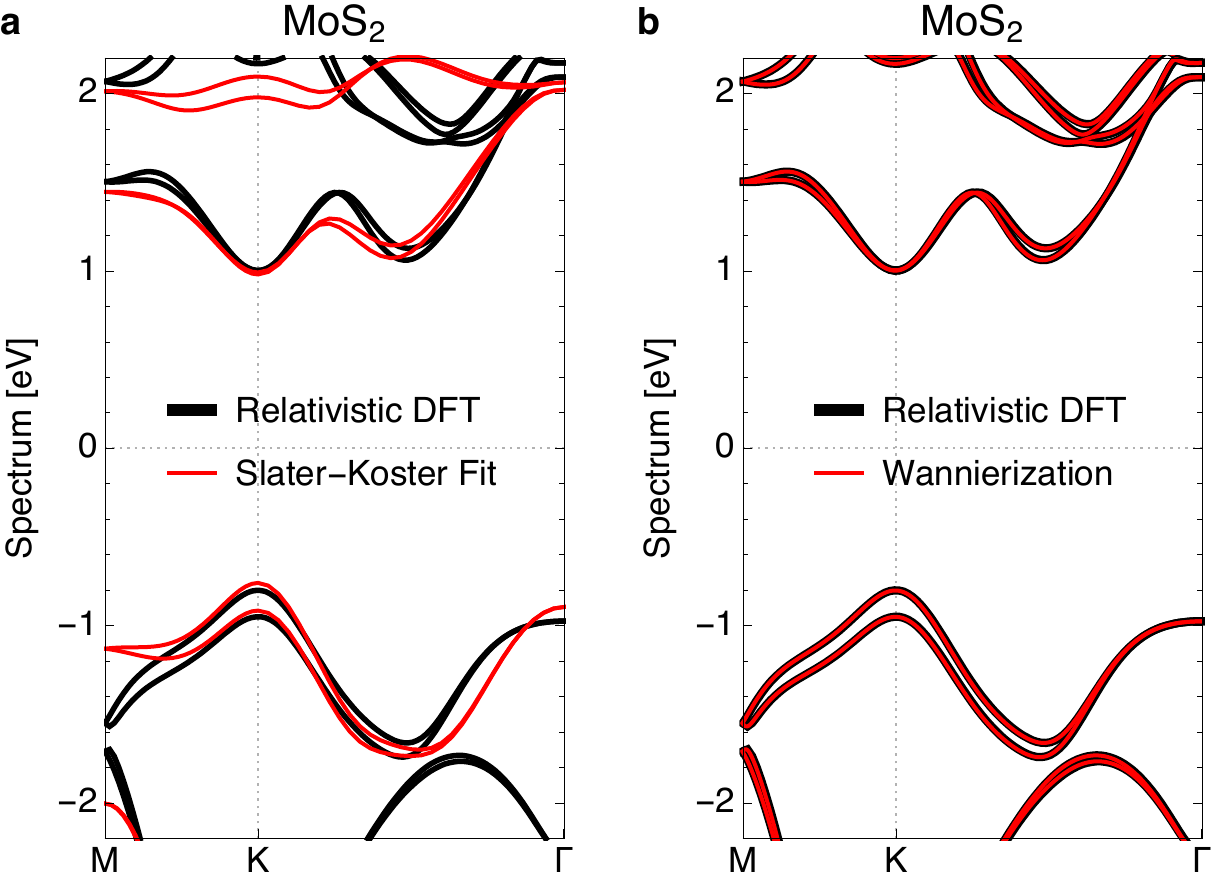} 
   \caption{MoS${}_2$ bandstructure comparison using tight-binding (red) and relativistic Density Functional Theory (black). Panel (a) shows a fit of a Slater-Koster model up to next-nearest-neighbors, adapted with permission from Ref. \cite{Silva-Guillen:AS16}. Panel (b) shows a Wannierization calculation with hoppings up to sixth-order neighbors, adapted with permission from Ref. \cite{Lado:2M16}.}
   \label{fig:TB}
\end{figure}

For larger-scale nanostructures and devices, inaccesible to \emph{ab-initio} approaches, the method of choice is typically tight-binding (TB), which occupies an intermediate level of computational complexity. The TB models for these systems can become moderately complicated, and may include many different crystal fields (onsite energies) and inter-orbital overlaps (hopping amplitudes) beyond nearest neighbors. All these parameters are often tuned to match the electronic structure of \emph{ab-initio} calculations. There are two main ways to do this. On the one hand, one may choose a set of relevant atomic orbitals in the unit cell of the material, fix the range of hoppings to consider and derive a general Slater-Koster model for the system, parametrized by a set of overlap integrals and crystal fields.\cite{Slater_PR_1954} These can then be fitted to match the \emph{ab-initio} bandstructure as closely as possible, potentially taking into account also the orbital character of the bands at different $k$-points. Such approach produces a sensible Slater-Koster TB model that allows in particular to incorporate the effect of strain in a natural way, by modifying hopping amplitudes $t_{ij,0}^{\alpha\beta}$ between orbitals $\alpha$ and $\beta$ as
\begin{equation}\label{strain}
t_{ij}^{\alpha\beta}=t_{ij,0}^{\alpha\beta}\left(1-\Lambda^{\alpha\beta}\frac{|\vec r_{ij}-\vec r^0_{ij}|}{|\vec r^0_{ij}|}\right).
\end{equation}
Here $\vec r_{ij}$ and $\vec r^{(0)}_{ij}$ are the vectors connecting sites $i$ and $j$ under strain and at equilibrium, respectively. Dimensionless parameters $\Lambda^{\alpha\beta}$ may be adjusted to \emph{ab-initio} results (even making them depend on $i,j$), or relating them to the angular momentum of $\alpha$ and $\beta$ using TB theory (Harrison rule). However, the quality of the fit using the Slater-Koster approach is often not excellent. Also, the fit procedure may become very delicate given the large number of free parameters, and the result may depend on which parts of the bandstructure one chooses to emphasise. For a more systematic and deterministic way to build TB models it may be preferrable to employ `Wannierization' techniques instead. In essence, Wannierization consists on a projection of the DFT bandstructure onto maximally localized states in real space of a given orbital character. The result is a TB Hamiltonian that is not of a Slater-Koster form, but that, if allowed to range beyond nearest neighbors, can often yield a precise fit of both bands and orbital character in a controlled way. Figure \ref{fig:TB} shows a comparison of both methods in the case of MoS${}_2$, taken from Refs. \cite{Silva-Guillen:AS16,Lado:2M16}.

\begin{figure}
   \centering
   \includegraphics[width=0.8\columnwidth]{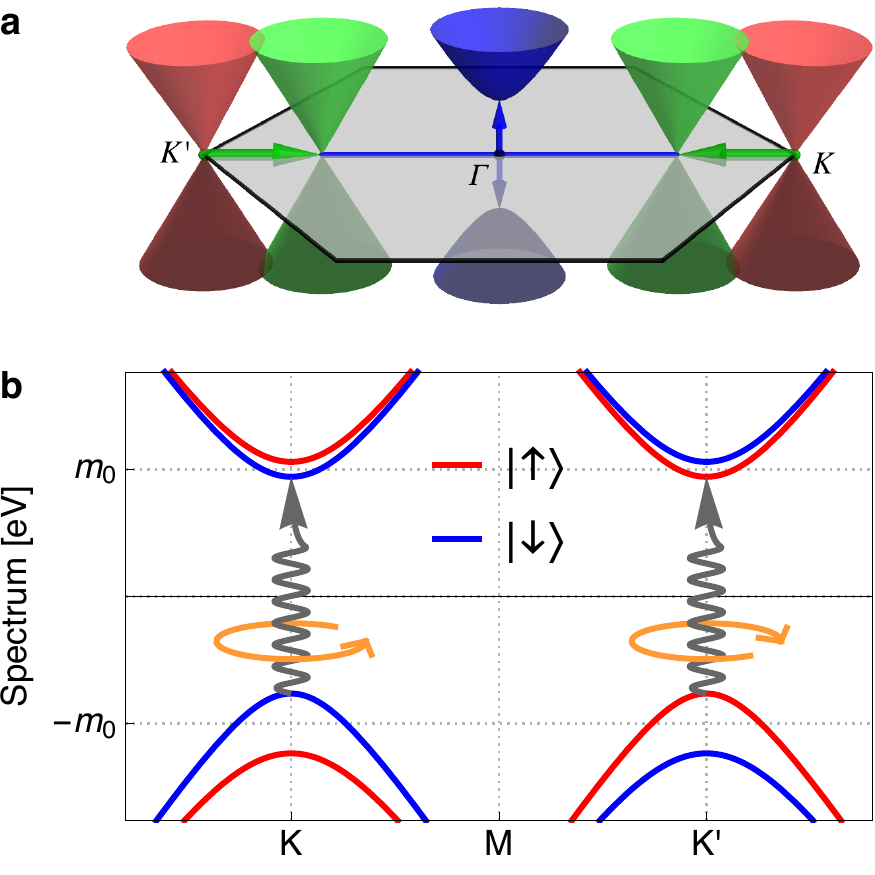} 
   \caption{(a) Evolution of graphene's Dirac cones (red) under uniaxial tension along the armchair direction (green). If the Dirac cones fuse at the $\Gamma$ point (blue), a gap opens, as in the case of phosphorene. (b) Transition metal dichalcogenides have gapped Dirac cones with a strong spin-orbit splitting of the valence band. Circularly polarized light may selectively excite electrons from a specific valley.}
   \label{fig:strain}
\end{figure}

The final level of the model hierarchy consists of analytical approaches. These are effective models useful to describe in a transparent way the essential physical mechanisms at play in a 2D crystal without aiming for a quantitative description. The relevant model for a given crystal depends on the specific space symmetries of its lattice. The most archetypal of these, relevant for honeycomb lattices (mono- and dichalcogenides, h-BN, graphene) is the massive Dirac equation in two dimensions, in effect a $\bf{k}\cdot\bf{p}$ extension of the effective mass approximation of semiconductors. Honeycomb crystals typically possess a bandstructure with a gap centred at two identical $\pm$ valleys located at the K and K' points in the BZ. Graphene is a special case, with gapless valleys, see red cones in Fig. \ref{fig:strain}a. The idea is to consider a neighborhood of the bandstructure around the two valleys, neglecting all but the  (spinful) valence and conduction subbands. Expanding these to second order in wavevector $\vec k=(k_x, k_y)$ around K and K' and imposing invariance under the $C_3$ symmetry group of the lattice ($120^\circ$ rotations), we arrive at a valley-degenerate Hamiltonian of the form $H=H_+\oplus H_{-}$, where each $\tau=\pm$ valley is described, in the absence of strains, by
\begin{equation}\label{Dirac}
H_\tau=\left(\begin{array}{cc}
m_0 +(\alpha+\beta)|\vec k|^2& \hbar v(\tau k_x-i k_y) \\
\hbar v(\tau k_x+i k_y) & -m_0 +(\alpha-\beta)|\vec k|^2
\end{array}\right) + H_\mathrm{so}
\end{equation}

The gap is $E_g=2m_0$, $v$ is a velocity and $\alpha$ and $\beta$ control the valence and conduction effective masses. The $H_\tau$ matrix above is expressed in the basis of $\vec k=0$ valence and conduction states, which define a `pseudospin'. The $\vec k=0$ conduction band state (`peudospin-up') becomes mixed with the $\vec k=0$ valence band state (`peudospin-down') due to the off-diagonal $\hbar v(\tau k_x \pm i k_y)$ terms, which makes pseudospin a non-trivial degree of freedom in any scattering (i.e. $\vec k$-changing) process. The pseudospin should not be confused with the real electron spin $s_z$ that enters the SOC term $H_\mathrm{so}$, which reads
\begin{equation}\label{Hso}
H_\mathrm{so} = \tau s_z\left(\begin{array}{cc}
\frac{\lambda_0+\lambda}{2} +(\lambda'_0+\lambda')|\vec k|^2& 0 \\
0 & \frac{\lambda_0-\lambda}{2} +(\lambda'_0-\lambda')|\vec k|^2
\end{array}\right),
\end{equation}
for some $\lambda_0$, $\lambda_0'$, $\lambda$ and $\lambda'$.  This SOC polarizes the real spin in opposite directions in opposite valleys. 
All parameters in the model may be obtained from experiment, from a TB or from \emph{ab-initio} models. States within each valley are characterized by an opposite angular momentum associated to the circulation of the pseudospin around the K and K' points in the BZ. This follows from the form of $H_\tau$, which has a non-zero `Berry curvature',\cite{Xu_NP_2014} and leads to valley selectivity of various properties, such as transport or the absorption of circularly polarised light, to be discussed below.

The effect of a strain field $\epsilon_{ij}$ at low energies can also be derived analytically. \cite{Rostami:PRB15} Its dominant contribution enters as a valley shift in graphene $\vec k\to \vec k - \eta\tau\vec A$ or as a deformation potential $H_\tau \to H_\tau + \eta'\mathcal{D}$ in other gapped honeycomb crystals. Here, $\vec A=(\epsilon_{xx}-\epsilon_{yy},-2\epsilon_{xy})$ is a pseudo-gauge vector field, expressed in terms of the strain tensor components, $\mathcal{D}\approx \epsilon_{xx}+\epsilon_{yy} +\mathcal{O}(\epsilon^2_{ij})$ is a scalar and $\eta$ and  $\eta^{\prime}$ are parameters that control the strength of the electromechanical coupling.

\subsection{Graphene: massless Dirac}

Graphene, the carbon honeycomb 2D crystal and forefather of all other 2D materials, corresponds to a particularly simple instance of the model in Eq. \eqref{Dirac}, as a result of its enhanced lattice symmetry group $C_{3v}$. The two orbitals that form the pseudospin in the Dirac model are $\pi$ orbitals of carbon atoms arranged into two triangular sublattices. Unlike in most other honeycomb 2D crystals, these two orbitals are indistinguishable in graphene, so that in addition to the $120^\circ$ rotations of $C_3$, the lattice is also invariant under inversion $\vec r_i\to -\vec r_i$ of all its site positions. The implications are profound. In particular the model cannot contain terms that distinguish the two sublattices, so $m_0=0$, and the system must be gapless. This argument points to a connection between the size of the gap and the chemical imbalance between the two sublattices in honeycomb crystals.  

Graphene's low energy spectrum is thus composed of two isotropic massless Dirac cones (the $\alpha,\beta$ corrections in (\ref{Dirac}) can be neglected for most purposes, and spin-orbit is also negligible due to the low atomic mass of Carbon). The massless Dirac spectrum exhibits pseudospin chirality and a unique scale invariance ($\vec r \to \Lambda \vec r$ and $\epsilon \to \epsilon/\Lambda$), responsible for many of graphene's remarkable electronic properties.  Amongst these are an absence of Anderson localization (provided disorder is smooth on the scale of the lattice spacing), the associated Klein tunneling phenomenon, or a logarithmic Fermi velocity renormalization from interactions.\cite{Castro-Neto_RMP_2009} 
One of the most remarkable consequences of graphene's scale invariance is the way strain couples to electrons. As mentioned above, a given (possibly position dependent) strain tensor $\epsilon_{ij}$ applied to the system enters the effective massless Dirac model as an effective pseudogauge field that shifts the position of the two Dirac cones in opposite directions, see green cones in Fig. \ref{fig:strain}a. 
If a sufficiently strong uniaxial strain is applied along the armchair direction, it can in principle have the dramatic effect of fusing the two Dirac cones at the $\Gamma$ point of the BZ. This possibility requires unrealistic deformations beyond graphene's point of rupture, but signals a possibility that is actually materialized in a different 2D crystal: phosphorene.

\subsection{Phosphorene: fused Dirac cones}

The lattice of black phosphorus monolayers is not of the honeycomb type, as its buckling along the armchair direction breaks the $C_3$ symmetry, see Fig. \ref{Fig:Bands}. Its gap is not located at two valleys.  In terms of symmetries, black phosphorus monolayers are analogous to graphene with a uniform uniaxial strain along the armchair direction, i.e. $C_3$ symmetry is broken but inversion is not. Moreover, spin-orbit coupling is also rather small. The TB model of the system is therefore much like graphene's, with the exception that of the three hopping amplitudes connecting a phosphorus $\pi$ orbital to its nearest neighbours, two are equal ($t_1<0$), but the other ($t_2>0$) is different and of opposite sign,\cite{Rudenko_PRB_2015} as would correspond to Eq. \eqref{strain} with an (unrealistically) large $|\vec r_{ij}-\vec r^0_{ij}|$ along the armchair direction. One may thus continuously transform the graphene TB model with $t_2=t_1<0$ into that of black phosphorus by increasing $t_2$, which introduces a pseudogauge-like shift of the two Dirac cones that approach each other. For $t_2<-2t_1$ the model remains gapless, but as $t_2\geq-2t_1$ the two cones fuse at the $\Gamma$ point. A gap $E_g \approx 2t_2+4t_1$ like phosphorene's then opens, abruptly changing the topology of the bandstructure, as depicted in Fig. \ref{fig:strain}a. 

This process can actually be reversed using strain. If we apply uniaxial compression to phosphorene along the armchair direction, $|t_1|$ is increased, while $t_2$ slightly decreases, so that the quasiparticle gap ($E_g\approx 1.8$ eV using DFT-GW at equilibrium) decreases (increases) by a huge $\sim 6$\% per 1\% of uniaxial compression (expansion). For a sufficient compression, we may reach $t_2\leq-2t_1$, which corresponds to a gapless spectrum. This situation may be achieved in practice for multilayers,\cite{Kim:S15} which then transition into a semimetalic phase with shifted Dirac cones. 

\subsection{Transition metal dichalcogenides}

The family of TMDs, of which MoS${}_2$ is probably the most studied, also share a honeycomb lattice structure. The difference between the metal Mo atom and the chalcogen S atom makes them semiconducting, with a spectral gap $E_g=2m_0$ of the order of 1.9 eV at the K and K' points. 
The response of the gap to uniaxial strain is opposite to that of phosphorene, decreasing by around 1.5\% per 1\% of uniaxial tension.

The large atomic mass of the metallic species endows these materials with a very sizeable spin-orbit coupling. This leads to a significant splitting of the valence bands (the splitting in the conduction band is quite smaller). 
The corresponding spin structure around the gap is sketched in Fig. \ref{fig:strain}b.  The opposite spin-polarization of opposite valleys opens unique opportunities for spintronics in these materials, as discussed below.

\section{Novel developments}\label{Sec:Novel}

\subsection{Control of spin and valley degrees of freedom}

Electron spin is an internal quantum degree of freedom associated with a magnetic moment, whose manipulation and control for electronic applications are studied in the field known as {\it spintronics}. The pseudospin and the valley index are additional degrees of freedom  that can play a role similar to spin in spintronics.\cite{San-Jose:PRL09} We have seen that the band structure of graphene and TMDs present two degenerate and inequivalent valleys at the K and K' points of the BZ.  While graphene is centrosymmetric, inversion symmetry is broken in single layer MoS$_2$ and related TMDs. The latter present a strong SOC, lifting spin degeneracy in both valence and conduction bands. Time reversal symmetry imposes a relation between valley and spin quantum numbers. Additionally, the opposite pseudospin angular momentum in each valley imposes selection rules for inter-band optical transitions under circularly polarized light, so that opposite light polarizations are absorbed  by opposite valleys, see Fig. \ref{fig:strain}b. The combination of both effects allows the use of optical helicity to control the spin polarization of photoexcited carriers at each valley independently. \cite{Xu_NP_2014} 
Since inter-valley scattering is highly suppressed due to the breaking of spin degeneracy, long spin and valley polarization lifetimes are possible in single layers of TMDs. A related phenomenon is the spin (valley) Hall effect, whereby opposite Hall currents for opposite spins (valleys) emerge. In single layer TMDs, the valley Hall effect can be realised in optoelectronic setups by using the optical valley-selection rules to create a population imbalance between different valleys. This produces a valley-polarized Hall current that switches sign with the light polarisation. Therefore, 2D TMDs are materials with potential to use the valley index in an equivalent way as the spin degree of freedom is used in spintronics, leading to the concept of {\it valleytronics}.\cite{Schaibley_NRM_2016}  Note that inversion symmetry and spin degeneracy at each valley are restored in bilayer and bulk TMD samples, demonstrating again the singular physics shown by some crystals when they are reduced to their monolayer form.

\subsection{Excitons}\label{Sec:Excitons}

\begin{figure}
   \centering
   \includegraphics[width=\columnwidth]{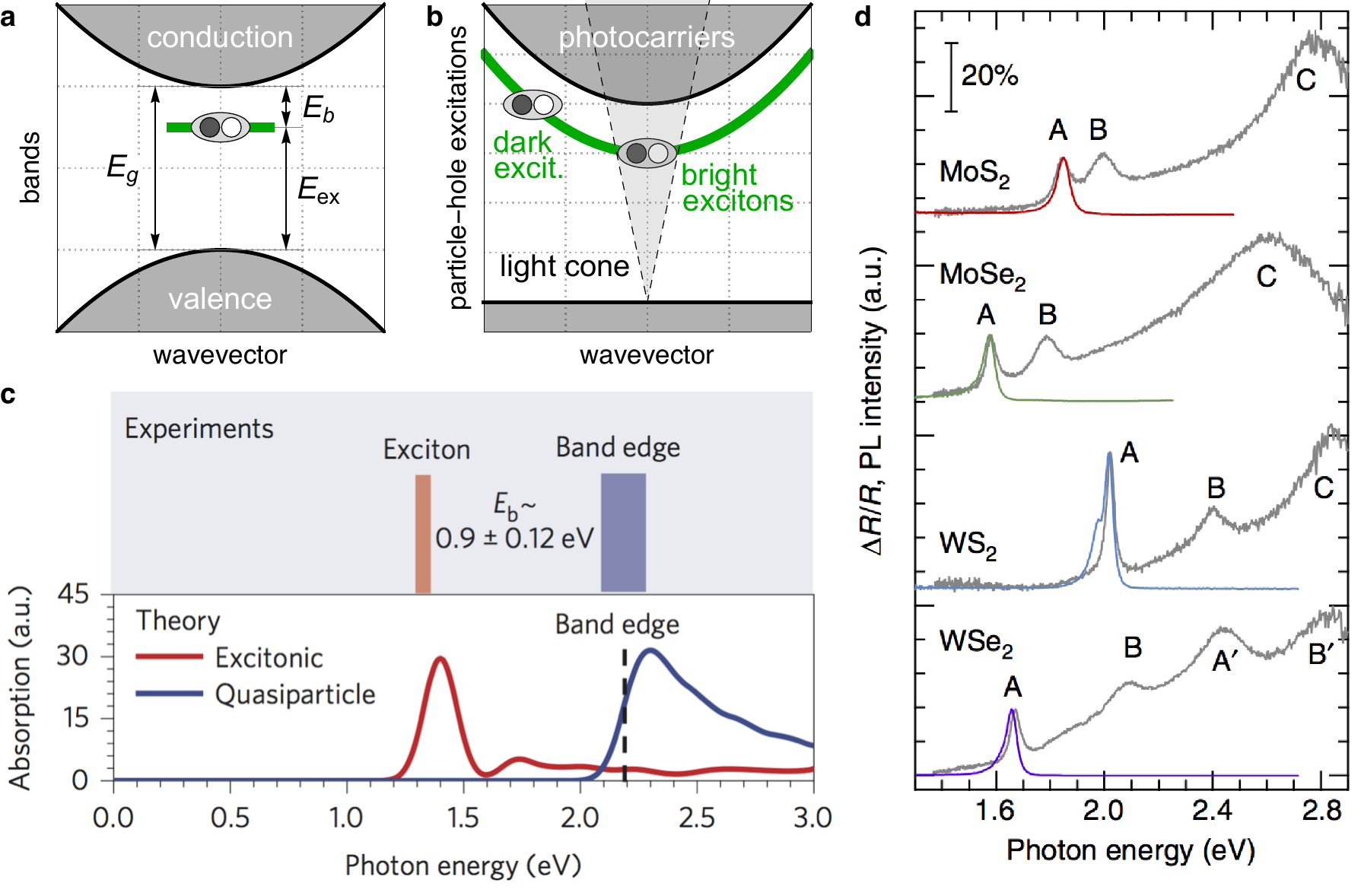}
   \caption{(a) Schematic representation of the single-particle energy bands, with an exciton state at energy $E_b$ below the gap $E_g$. (b) Schematic representation of the particle-hole excitation spectrum. Excitons within the light cone can decay radiatively and are thus "bright", see Ref. \cite{San-Jose_PRX_2016}. (c) Excitons in phosphorene from Ref. \cite{Wang:NN15}. Top: schematic plot showing the measured ground-state exciton energy (red) and the energy corresponding to the quasiparticle band edge (blue). Bottom: Calculated optical absorption from first-principles of monolayer black phosphorus with e-h interactions (excitonic absorption, red curve) and without e-h interactions (quasiparticle absorption, blue curve). (d) Photoluminescence spectra (red, green, blue and purple curves) and differential
reflectance spectra (grey curves) from Ref. \cite{Kozawa:NC14} for a variety of TMDs exhibiting a number of different excitonic peaks. 
 Adapted from Ref. \cite{San-Jose_PRX_2016} with permission of the American Physical Society [panel a)], from Refs. \cite{Wang:NN15} and \cite{Kozawa:NC14} with permission from Nature Publishing Group [panels b) and c)].}
   \label{fig:excitons}
\end{figure}

By shining materials with light and analysing their optical properties one can in turn acquire information about their electronic structure. When photons of sufficient energy hit a semiconductor, they can excite electrons from the valence band to the conduction band creating electron-hole (e-h) pairs. Within a one-electron approximation, these pairs can be described in terms of the material's band structure. An excited e-h pair then has a minimum energy equal to the bandgap $E_g$. However, since these quasiparticles are charged, they interact through the electrostatic Coulomb force and a proper description of the excitation spectrum has to take into account the many-body nature of their wave function. It turns out that the attraction between the excited electron with charge $-e$ and the hole left behind with charge $+e$, causes their motion to be correlated and decreases the energy of the pair with respect to the free e-h excitation, forming a bound entity with a minimum energy $E_\mathrm{ex}<E_g$ inside the gap that is called an \emph{exciton}, see Fig. \ref{fig:excitons}a. The exciton has a finite lifetime, as the electron and the hole can undergo recombination into the original unexcited ground state. An exciton is thus a metastable condensed matter elementary excitation that carries energy and momentum with zero net charge. There are other higher order excitations, such as trions, biexcitons, polaritons, etc. that we will not discuss here.

The typical quantities that describe an exciton are its binding energy, $E_b\equiv E_g-E_\mathrm{ex}$
and its size, parametrized by the exciton radius, $a$. In a few materials where the Coulomb interaction is very strong, as is the case of fullerenes, small excitons of a size comparable to the lattice constant may form, which are known as Frenkel excitons. These have binding energies of the order of 1 eV. In the most common cases, the semiconductor dielectric constant is large and electric field screening tends to reduce the Coulomb interaction, producing bigger exciton radii, smaller binding energies of the order of $10 - 100$ meV, and longer exciton lifetimes. These are called Wannier-Mott excitons.

To find the exciton solution exactly is certainly complicated, but can in principle be done by solving the so-called Bethe-Salpeter equation for excitons. In many cases, though, we can perform a series of simplifications valid for most common semiconductors. First, we consider that the electron that is excited into the conduction band can be simply described as a quasiparticle with mass $m_e$ as obtained by the conduction band minimum. Its interaction with the rest of the electrons of the valence band is replaced by its interaction with a hole with opposite charge and mass $m_h$ given by the valence band structure at its maximum. This is known as the effective mass approximation. Second, we consider that different e-h pairs are so far apart (in space or in time), that they can be considered independent. These two assumptions allow us to neglect exchange and correlation contributions to the Coulomb interaction. The two-particle problem is then solved by going to the center of mass and relative coordinate reference system. 

The wave function of the resulting bound state is said to be hydrogenic because it is similar to the one of the hydrogen atom.
However, the Coulomb interaction in a typical 3D semiconductor $V_C(r)=-e^2/(4\pi\epsilon_0\epsilon r)$
is screened by all other electrons in the valence band. This effect is captured by a dielectric constant $\epsilon\gg 1$.
The effective masses $m_e$ and $m_h$ are, moreover, much smaller than the free electron mass $m$. This results in an exciton binding energy $E_b$ much smaller than the Rydberg energy, and an exciton radius $a$ much bigger than the hydrogen atom,\cite{Yu:05}
\begin{equation}
E^{\rm{3D}}_b=\frac{\mu}{m\epsilon^2}R_{\infty},\,\,\,\,\,\, a^{\rm{3D}}=\frac{m\epsilon}{\mu}a_0.
\end{equation}
Here $R_{\infty}\equiv \frac{e^2}{4\pi \epsilon_0}\frac{1}{2 a_0}=13.6$ eV is the Rydberg energy, $\mu=m_e m_h/(m_e+m_h)$ is the exciton reduced mass and $a_0=4\pi\epsilon_0\hbar^2/m e^2=0.53\,\textrm{\AA}$  is the Bohr radius.

Now, in a 2D crystal, the electron and hole are confined to move in lower dimensions, thus increasing their attraction. Constrained to 2D space, the same Coulomb interaction $V_C(r)$ translates into a bigger $E^{\rm{2D}}_b=4E^{\rm{3D}}_b$ and a smaller $a^{\rm{2D}}=a^{\rm{3D}}/2$. This is a rather rough simplification of the problem, however, that neglects the different screening between the 2D material and its surrounding environment.
Keldysh showed that in thin films of thickness $d$, the actual interaction $V_K(r)$ between charges at distances bigger than $d$ is influenced by the the dielectric constant of the medium surrounding the film, $\epsilon_1$ and $\epsilon_2$ (typically the substrate and vacuum). Solving the electrostatic problem with the use of image charges,\cite{Keldysh:JL79} he obtained  $V_K(r)=R_\infty(\pi a_0/\bar{\epsilon}r_0)\left[Y_0\left(r/r_0\right)-H_0\left(r/r_0\right)\right]$, where $\bar{\epsilon}=(\epsilon_1+\epsilon_2)/2$, $r_0=d\epsilon/(\epsilon_1+\epsilon_2)$ is an effective screening length and $Y_0$ and $H_0$ are second-kind Bessel and Struve functions, respectively. This potential presents a logarithmic divergence for $r\rightarrow 0$ (like the potential of a charged string) and reduces to the unscreened Coulomb potential at large distances,
\begin{equation}
V_K(r)
\approx\frac{2 a_0 R_\infty}{\bar{\epsilon}r_0}\left\{\begin{array}{ll}
\gamma+\log\left(\frac{r}{2r_0}\right) & r\ll r_0\\
-r_0/r & r\gg r_0
\end{array}\right.,
\end{equation}
where $\gamma$ is Euler's constant. The crossover between these two behaviours is characterised by the length scale $r_0$. Cudazzo et al. \cite{Cudazzo:PRB11} later considered the problem of strictly $d=0$ 2D crystals (embedded in a dielectric environment) and arrived at the same Keldysh potential, but expressed in terms of the 2D polarizability $\alpha_{\rm{2D}}$ of the dielectric sheet. The polarizability determines the scale at which the two asymptotic forms match, $r_0=2\pi\alpha_{\rm{2D}}$.
 
For excitons larger than $r_0$, the binding energy and the exciton radius have the same form as $E^{\rm{2D}}_b$ and $a^{\rm{2D}}$, but with $\epsilon$ replaced by $\bar\epsilon$ (note that $\bar\epsilon=1$ for the 2D dielectric in vacuum). This approximation is good for excitons whose size is large as compared to $r_0$. In the opposite limit of small excitons, the e-h pair experiences mostly the logarithmic part of the Keldysh interaction, and\cite{Prada:PRB15} 
\begin{equation}
E^{\rm{2D}}_b=\frac{a_0 R_\infty }{\bar{\epsilon}r_0}\left[2 \log\left(\frac{4 r_0}{a^{\rm{2D}}}\right)-3\right], \,\,\,\,\,\, a^{\rm{2D}}=\sqrt{\frac{\bar{\epsilon}m}{\mu}a_0 r_0}.
\end{equation}
This is the case of monolayer MoS$_2$ or phosphorene, with binding energies around 600 meV,\cite{Castellanos-Gomez:2M14,Prada:PRB15,San-Jose_PRX_2016} mid-way between the typical Wannier-Mott and Frenkel scales.

The excitons influence the optical properties of semiconductors and can be detected in experiments. One such property is the optical absorption, which is the conversion of a photon into an exciton. Due to conservation of energy and momentum, this process occurs at points in momentum space where the photon light cone overlaps with the exciton dispersion surface, see Fig \ref{fig:excitons}b. The existence of excitons lowers the threshold of photon absorption, which shows peaks at the energies of different internal states of the exciton, see Fig. \ref{fig:excitons}c for the case of phosphorene. Said absorbance may be measured through the differential reflectance $\Delta R/R$, \cite{Kozawa:NC14} which thus exhibits signatures from different internal exciton states that may decay radiatively or non-radiatively, see Fig. \ref{fig:excitons}d. 
It is also possible to measure the opposite phenomenon, the material's photoluminescence (PL). In PL, the electrons in the sample are excited electrically or optically and, after some energy loss (relaxation), they recombine and return to the ground state radiatively, i.e. by emitting light. Fig. \ref{fig:excitons}d shows different PL spectra for a series of TMDs.

\subsection{Strain engineering}

\begin{figure}
\centering
  \includegraphics[height=7cm]{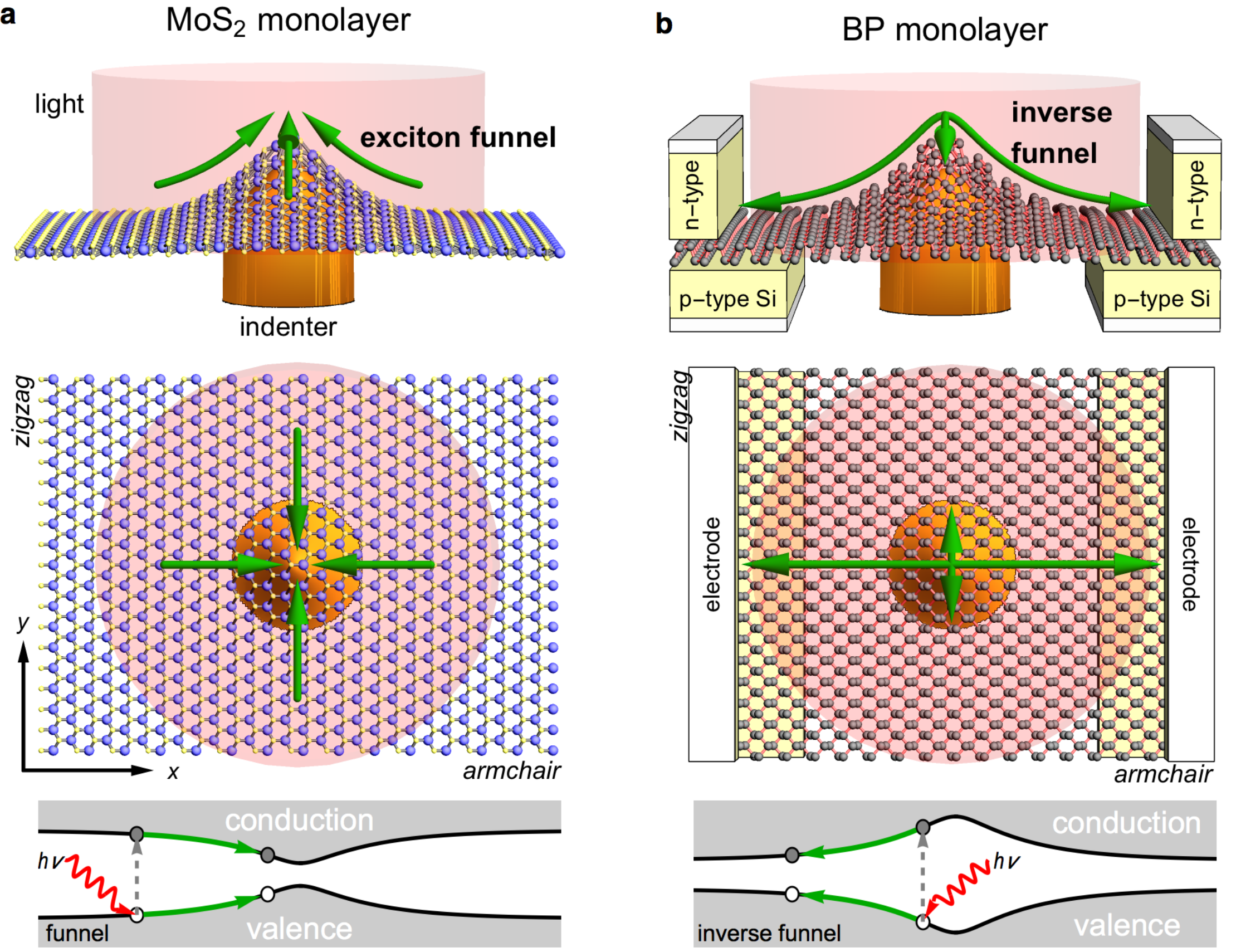}
  \caption{Strained induced funnel effect of excitons in 2D crystals. An indenter creates an inhomogeneous profile of strain in single layer MoS$_2$ (a) and phosphorene (b). The strain gradient modulates the gap of the two crystals as sketched in the bottom panel. Photogenerated excitons (green arrows) are pushed isotropically towards the indenter center in MoS$_2$ (funnel effect), while they are pushed anisotropically away it in black phosphorus (inverse funnel effect). Adapted from Ref. \cite{San-Jose_PRX_2016} with permission of the American Physical Society.}
  \label{Fig:Funnel}
\end{figure}

The coexistence of high stiffness and flexibility in 2D crystals  make these materials excellent platforms for {\it strain engineering}, i.e. applying strain externally to tune and control the electronic and optical properties of materials.\cite{Roldan_JPCM_2015} Some possibilities are:

{\bf Generation of pseudo-magnetic fields--} As discussed in Sec. \ref{Sec:Models}, when graphene is subjected to strain fields, the modification of hopping amplitudes between carbon atoms gives rise to effective gauge fields, whose effect is similar to that of a magnetic field applied perpendicular to the graphene plane, albeit opposite in different valleys.\cite{Guinea_NP_2010} 
Therefore, strain engineering can be used to discretize the graphene band structure into a set of pseudo-Landau levels, corresponding to counterpropagating cyclotron orbits in opposite valleys.
This effect has been observed experimentally and pseudo-magnetic fields exceeding 300 Tesla have been reported in trigonally distorted graphene nanobubbles.\cite{Levy_S_2010}  MoS$_2$ and related TMDs have, like graphene, an hexagonal lattice structure. However, the simple pseudo-magnetic field picture of strained graphene does not carry over to deformed monolayer TMDs. Due to the complex orbital contributions leading to the formation of the valence and conduction bands, we have seen in Sec. \ref{Sec:Models} that the effect of strains in the low energy model is dominated by different terms than in the massless Dirac Hamiltonian relevant for graphene. As a consequence, not only one gauge field but several pseudo-vector potentials and scalar fields appear in strained single layer TMDs.\cite{Rostami:PRB15} The inclusion of strain-displacement relations from valence force-field models leads to additional correction to the strain-induced pseudo-magnetic fields.\cite{Midtvedt_2DM_2016}

{\bf Direct-to-indirect gap and semiconducting-to-metallic transitions--} The outstanding stretchability of 2D semicondcutors can be used to tune the size and the nature of their band gap. While single layer TMDs are direct gap semiconductors, uniaxial strain can produce a shift of the band edges and drive a transition to an indirect gap. The opposite trend takes place in group IV-B monochalcogenides, which are indirect gap semiconductors that can become direct gap crystals under tensile strain. Controlling the nature of the gap with strain ({\it straintronics}) may become a powerful strategy for photonics applications. One may envision combining dark and bright regions for photoexcitation within the same sample by tailoring substrate-induced tensions. Furthermore, by applying higher amounts of strain, but still below the fracture limit, it is possible to drive a semiconducting-to-metallic transition. Such a huge modification of the gap in 2D semiconductors (from $\sim 1.9$~eV to 0~eV for the case of MoS$_2$) has to be compared with the poor tunability of 3D semiconductors like silicon of only $\sim 0.25$~eV under $\sim 1.5\%$ of biaxial strain.

{\bf Strain induced funnel of excitons--} Strain engineering can be used for the creation of a broad-band optical funnel for excitons in semiconducting 2D crystals.\cite{Feng_NP_2012} By continuously changing the strain across a sheet of MoS$_2$ or black phosphorus, for example, a continuous variation of the optical band gap is produced, allowing for the capture of photons with different energies. Furthermore, it is possible to tune the strain profile so as to drive the photogenerated excitons towards the regions of minimal gap, creating a {\it funnel} for excitons.\cite{Castellanos-Gomez_NL_2013} The funnel effect in BP is much stronger than in MoS$_2$ and of opposite sign.\cite{San-Jose_PRX_2016} While excitons in MoS$_2$ are driven isotropically towards regions of maximum tension, excitons in BP move away from tensile regions. This difference stems from the fact that MoS$_2$ and BP respond differently to the application of strain: the gap in MoS$_2$ (BP) is reduced (increased) with tensile strain.
Furthermore, the exciton drift in MoS$_2$ is isotropic, while the funnelling in BP is highly anisotropic, with much larger drift lengths along one crystallographic (armchair) direction (see Fig. \ref{Fig:Funnel}). The inverse funnel effect can be beneficial for manipulation and harvesting of light, and in particular for the design of more efficient  solar cells.

{\bf Piezoelectricity--} A large number of 2D crystals lack inversion symmetry. This gives them piezoelectric properties, that is the potential ability to convert mechanical to electric energy. Stretching or compressing a piezoelectric crystal generates an electrical voltage. Conversely, an applied voltage produces expansion or contraction of the crystal. MoS$_2$, which is centrosymmetric in its 3D bulk configuration, has been shown to become piezoelectric when it is reduced to its monolayer form.\cite{Wu_N_2014} Recently, single layer monochalcogenides have been predicted to show an anomalously strong piezoelectric response, with piezoelectric coefficients that can be two order of magnitude larger than those in MoS$_2$. Furthermore, due to their anisotropic puckered lattice, the piezoelectric properties are strongly angle-dependent.\cite{Tian_NT_2016}  Considering the rapid advances in nanofabrication techniques, their amazing elastic properties and the possibility to withstand large strains, 2D piezoelectric crystals are viewed as promising platforms for applications in nano-sensors or portable electronic devices for energy harvesting.    

\section{Superconductivity}

\begin{figure}
  \centering
  \includegraphics[width=\columnwidth]{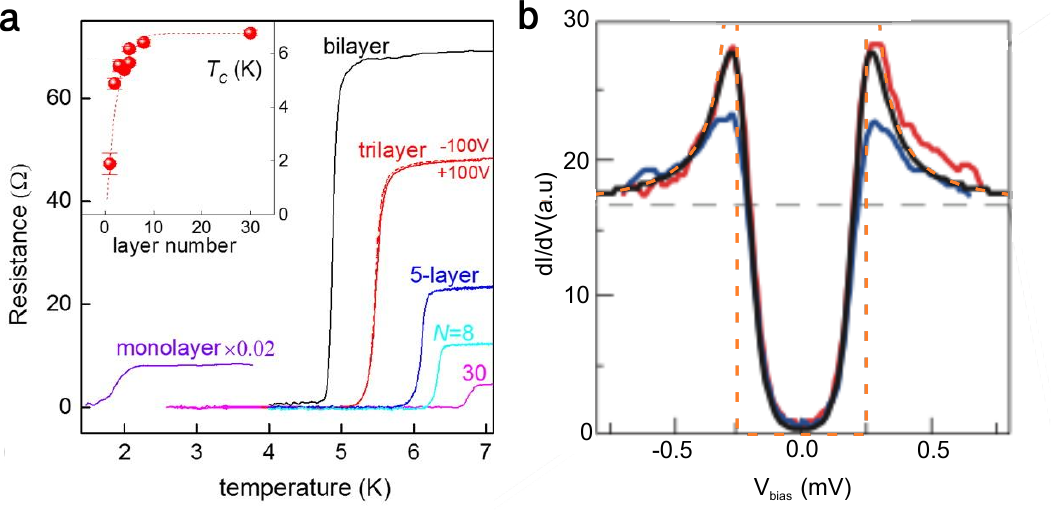}
  \caption{(a) Experimental variation of the resistance vs. temperature for different number of layers of NbSe$_2$. Shown in the inset is the dependence of $T_c$ with the number of layers. Adapted with permission from Ref. \citep{Cao} (b) Experimental tunnelling differential conductance dI/dV of a superconducting monolayer of Pb for different  representative tip positions (red and blue solid lines refer to regions of high and low coherence peak, respectively, the and the black solid line is the BCS best fit).  The dashed orange line corresponds to the theoretical  zero-temperature dI/dV spectra of a BCS superconductor, proportional to the density of states (DoS). The DoS is zero within the gap $\Delta$ and exhibits pronounced coherence peaks at $\pm \Delta$. Adapted with permission from Ref. \cite{Brun2014}. \label{fig:sc1}}
\end{figure}

Superconductivity is one of the most fascinating property of matter of pure quantum mechanical origin. It is well known that, by applying a current through 
a metal, a voltage drop across the sample is usually generated that is proportional to the applied current via the well known Ohmic law, 
$V=RI$, with $R$ the resistance. In ordinary metals, the resistance diminishes as  temperature is decreased, but eventually saturates to a finite 
value that is due to electrons scattering off impurities and imperfections of the sample.  In a superconductor, in contrast, an abrupt drop to zero resistivity occurs at a certain critical temperature, $T_c$, below which current flows through the system without resistance (see Fig. \ref{fig:sc1}a). 
A finite resistance generates dissipation, but in superconductors supercurrents can flow indefinitely without dissipating. The local density of states measured in Scanning Tunneling Spectroscopy reveals that a superconductor is characterized
 by a full gap $\Delta$ in the quasiparticle spectrum, with highly pronounced 
coherence peaks at energy $\pm\Delta$ (see Fig. \ref{fig:sc1}b). The absence of single-particle states at energies below the gap and the 
presence of supercurrents point to a non-trivial state of matter, where charge carriers are different from the conventional elementary quasiparticles. 

The celebrated Bardeen-Cooper-Schrieffer (BCS) theory of superconductivity (SC) explains the phenomenon as a many body phase-coherent state, where electrons with opposite spin and momentum pair up across the Fermi surface via a phonon-mediated attractive 
interaction. This pairing leads to the formation of new elementary entities, the so-called Cooper pairs, that Bose-condense to form a new ground state with a finite excitation gap $\Delta$ that reflects the binding energy of Cooper pairs. The BCS theory was a tremendous success, and was able to explain and predict all the essential properties of many 3D superconductors.

Thermal fluctuations break Cooper pairs apart. This makes the gap temperature-dependent, with $\Delta(T=0)=\Delta_0$ and $\Delta(T_c)=0$. For $T<T_c$, the coherence length $\xi(T)\propto 1/\Delta(T)$ measures the size of the Cooper pairs. At $T=T_c$, $\xi\to\infty$ and the Cooper pairs unbind.
According to Anderson theorem, $T_c$ for $s$-wave SC  is not 
affected by weak conventional disorder due to time-reversal (TR) invariance. A magnetic field breaks TR in a way that makes states of opposite spin 
and momentum on the Fermi surface no longer degenerate. For low fields the resulting non-dissipative Cooper pair motion screens the external 
field completely. The magnetic flux is thus completely expelled from the superconductor, which behaves as a perfect diamagnet. 
When the magnetic field is increased beyond a critical value, $H_{\rm cr}$, SC is destroyed. The value of $H_{\rm cr}$ corresponds to a magnetic length for Cooper pairs equal to their size $\xi$, $H_{\rm cr}=\Phi_0/(2\pi \xi^2)$, with $\Phi_0=h/2e$ the superconducting flux quantum.

When lowering the dimensionality of the system, these well-defined results must be revised. The density of states (DoS) at 
the Fermi level decreases, and so does the density of electrons available for pairing, and consequently $T_c$. In a 2D material with $N$ layers, the 
BCS theory predicts that $T_c(N)=T_c\exp\left(-1/(U\rho_0 N)\right)$ (Cooper law), with $\rho_0$ the single layer DoS at the Fermi level, and $U$ the pairing 
interaction strength. For a quasi-2D slab of thickness $d\ll \xi$, the in-plane critical field changes to $H_{{\rm cr},\parallel}\propto \Phi_0/(2\pi \xi d)$, 
so that the system may support much higher in-plane fields than the 3D case. In the extreme 2D limit $d\to0$, $H_{{\rm cr},\parallel}$ formally diverges. 
In this case, a second critical field appears, that is related to the pair-breaking action of the applied field via Zeeman spin polarization. Thus, the critical 
field in a strictly 2D BCS superconductor is the Pauli paramagnetic limit, $H_{{\rm cr},\parallel}=H_p\equiv\Delta/(\sqrt{2}\mu_B)$, with $\mu_B$ the 
Bohr magneton. Above the Pauli field $H_p$, the Zeeman splitting of the Cooper pairs compensates the energy gained from creating the BCS condensate, 
and 2D SC is suppressed.

The Bose condensation of Cooper pairs into the same state implies that the number of pairs (and hence of electrons) ceases to be well-defined. It follows 
that the BCS ground state is a superposition of states with different number of particles. The quantum uncertainty on the number of particles fixes the 
phase of the BCS ground state. It is thus said that SC spontaneously breaks the global gauge symmetry of the system, and develops a complex order parameter $\Delta({\bf r})=\Delta_0 e^{i\phi({\bf r})}$. In this language, the spectral gap is given by the average $|\langle\Delta\rangle|$. Thermal fluctuations of $\phi$ are responsible for the suppression of the gap with temperature. In a 3D superconductor below $T_c$, the phase $\phi({\bf r})$ fluctuates around a well-defined average that is fixed across the entire sample at equilibrium, so that the gap is finite and SC remains stable against fluctuations.

In 2D systems a dramatic change takes place. Thermal fluctuations in this case have a stronger effect, and destroy the long-range rigidity of the phase across the system.
The non-local correlation function 
$\langle \Delta^*({\bf r})\Delta({\bf r}')\rangle=\Delta_0^2\langle e^{-i\phi({\bf r})}e^{i\phi({\bf r}')}\rangle$, which remains finite at long distances $|\mathbf{r}'-\mathbf{r}|\gg \xi$ in 3D, is in contrast suppressed in 2D, reflecting a lack of long-range superconducting order,  
\begin{equation}
\langle \Delta^*({\bf r})\Delta({\bf r}')\rangle=\Delta_0^2 \left(\frac{|{\bf r}-{\bf r}'|}{r_0}\right)^{-\eta}\, \, \textrm{in 2D}.
\end{equation}
Here, $\eta(T)>0$ is a so-called critical exponent and $r_0$ a cutoff length scale of the order of the coherence length $\xi$. The superconductor
correlation function of the gap decays as a {\it power-law} of distance in 2D. This is a reflection of the Mermin-Wagner theorem, that states that 
there cannot exist long-range order in a finite-temperature 2D system by spontaneously breaking a continuous symmetry. The above power-law decay is still much slower than an ordinary exponential, so we may talk about so-called quasi-long-range order in 2D superconductors. In fact, 2D SC may survive in the form of a Berezinskii-Kosterlitz-Thouless (BKT) phase. This phase is a very complex state of matter that will not be discussed here.

The fundamental arguments sketched above suggest that 2D systems should be host to particularly non-trivial superconducting phases.
Superconducting thin films have been the subject of intense investigation during the last decades of the past century and most of their basic superconducting properties have been unveiled. It has been found that, as their thickness is reduced, thin films usually exhibit disordered structures, mostly amorphous and granular, that do not favor SC, or that show unusual behaviours, such as strong spatial fluctuations of the  quasiparticle peaks (see Fig. \ref{fig:sc1} b). In contrast, 2D crystals fabricated using methods such as exfoliation are very pure and clean down to atomic scales. They thus allow us to pursue 2D superconductivity into new territory. The newly emerging field of 2D superconductors\cite{SaitoRev,BrunRev} is now actively exploring the fundamentally new physics in these systems and their possible applications. We now present an overview of some of the recent experimental achievements in this domain with simple theoretical explanations.

\subsection{Quantum Confinement}

According to BCS theory, finding a quasi 2D superconducting system at relatively high temperatures would be very unlikely.
Experiments have nonetheless demonstrated that superconductivity persists when reducing the dimensionality of the samples. 
In particular, the controlled exfoliation of van der Waals materials like TMDs has made it possible to study the evolution of the 
superconducting properties of layered systems with the number of layers. In NbSe$_2$ a clear trend is reported,\cite{Cao} 
whereby $T_c$ decreases monotonically upon reducing the number of layers, with a law that approximately follows the BCS 
predictions. SC has been eventually reported down to the single layer limit (see Fig. \ref{fig:sc1}a). However, deviations from the Cooper law 
have been reported.

~~{\bf $T_c$ vs thickness --} 
Besides an overall decrease of $T_c$ with sample thickness, experiments in thin slabs of Pb \cite{BrunRev} have revealed 
oscillations of $T_c$ with the number of layers.  This  phenomenon is explained as follows. As the thickness of a film is reduced 
to the nanometer scale, the film surface and interface confine the motion of the electrons, leading to the formation of discrete 
electronic states known as quantum well states. This quantum size effect changes the overall electronic structure of the film and 
determines oscillations of $T_c$. A much more dramatic deviation from Cooper law has been reported in a TMD layered material, 
TaS$_2$, for which $T_c$ shows an enhancement upon reducing the number of layers,\cite{Navarro} contrary to the behavior of 
NbSe$_2$. These findings point to an interplay between dimensionality, strong Coulomb repulsion and existence of van Hove 
singularities in the DoS (i.e. logarithmic divergences in the DoS associated to saddle points in the band structure of 2D crystals).

{\bf Layered SC --}
Despite many theoretical predictions, superconductivity in graphene has long remained elusive. At the Dirac point the DoS is zero, 
so that SC is not expected at charge neutrality. The high Fermi velocity  at the Fermi level also requires strong doping in order to achieve a sizeable 
DoS and electron density. Superconductivity has been recently reported in graphene decorated with alkali metals,\cite{Ludbrook, Chapman} with the dopants 
intercalated between effectively decoupled graphene layers.
The dopants increase the electronic concentration and enhance the electron-phonon coupling. The $T$-dependent $H_{{\rm cr},\parallel}(T)$ exhibits 
a positive curvature that is consistent with the behavior of superconductors made of weakly coupled superconducting layers. 
This system realises layered SC, in which the vortex lines move between the layers, allowing the individual layers to remain superconducting at much higher fields.

\subsection{Unconventional properties}

Low dimensionality also allows for the emergence of novel SC effects associated to the shape of the Fermi surface (which is often nested), the reduction of symmetry, the appearance of van Hove singularities and the enhancement 
of Coulomb interactions, see Sec. \ref{Sec:Excitons}. 
One such example is the emergence of Ising superconductivity in system characterized by broken inversion symmetry and strong spin-orbit coupling, such as TMDs monolayers.

\begin{figure}
  \centering
  \includegraphics[width=0.75\columnwidth]{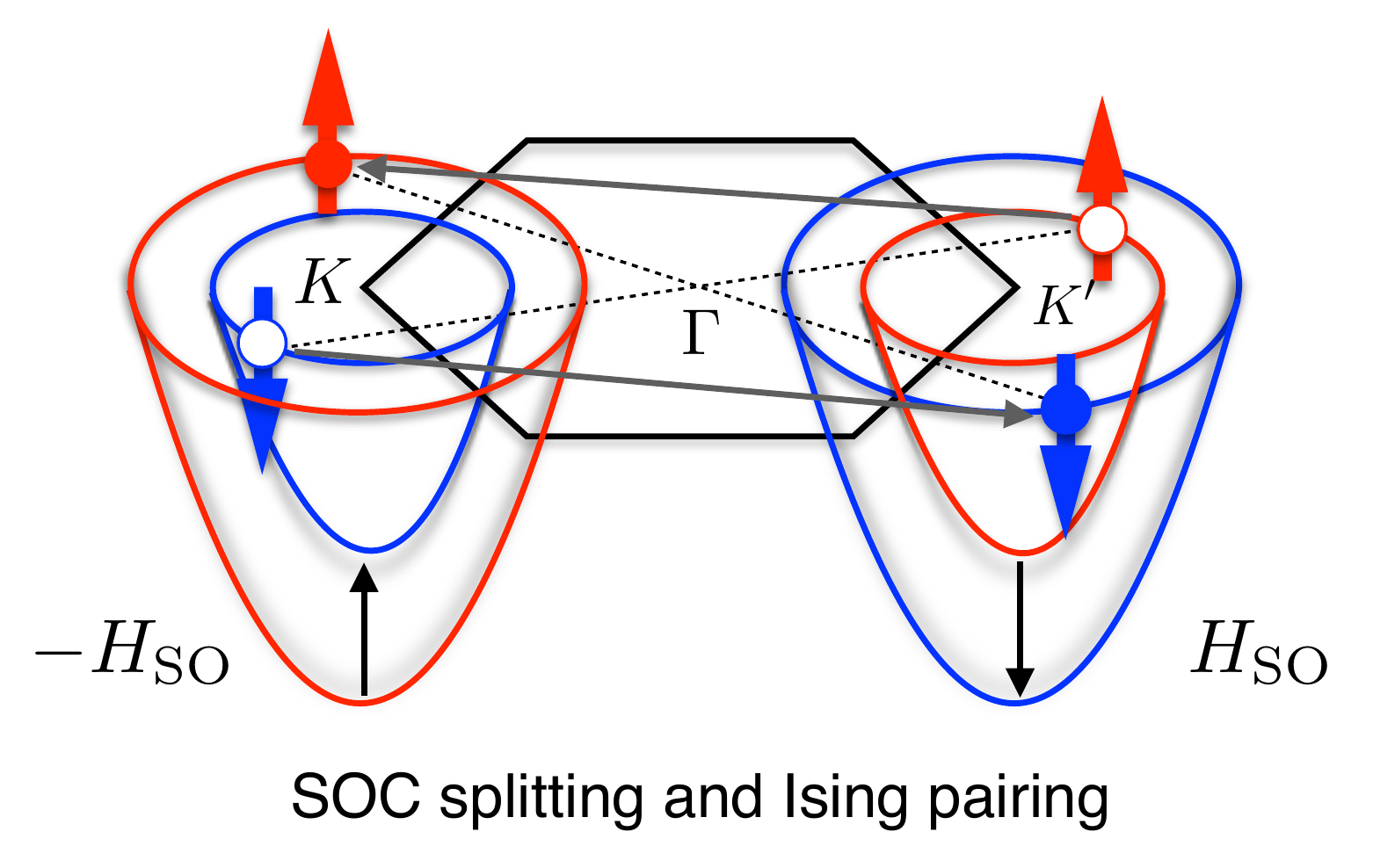}
  \caption{Schematics of the Ising pairing on the spin-orbit split Fermi surface of TMDs.}
  \label{fig:sc2}
\end{figure}

~~{\bf Ising SC --}
We have seen that, for 2D crystals, the in-plane critical field is given by Pauli paramagnetic limit $H_p$. In MoS$_2$ 
and NbSe$_2$,\cite{SaitoRev} it has been shown experimentally that $H_{{\rm cr},||}$ greatly exceeds the expected value from 
the Pauli limit. This behaviour has been ascribed to the strong spin-orbit coupling that arises in monolayer 
TMDs, see Sec. \ref{Sec:description}, which produces a strong spin--valley locking. 
The Fermi surface of this materials is formed by pockets around the K and K' points (in the case of MoS$_2$ this is achieved upon doping). 
As discussed in Sec. \ref{Sec:Models}, the SOC produces an effective Zeeman field $H_{\rm SO}$ with opposite sign at the 
K and K' pockets (see Fig.~\ref{fig:sc2}). The SOC spin-splits the Fermi pockets and 
polarizes the spin along the out-of-plane direction. Singlet superconductivity pairs electron of opposite spin across the Fermi 
surface, and Cooper pairs may form either as $|\textrm{K},\uparrow;\textrm{K'},\downarrow\rangle-|\textrm{K'},\downarrow;\textrm{K},\uparrow\rangle$ or as 
$|\textrm{K},\downarrow;\textrm{K'},\uparrow\rangle-|\textrm{K'},\uparrow;\textrm{K},\downarrow\rangle$, 
resulting in a so-called Ising pairing, as schematically depicted in Fig.~\ref{fig:sc2}. An external in-plane field $H_\parallel$ 
tends to tilt the spin towards the plane, in competition with the out-of-plane effective SOC field $H_{\rm SO}$, 
that tends to keep the spin aligned to the out-of-plane direction. This allows greater fields to be applied before destroying SC. The in-plane critical field may be estimated by noting 
that the in-plane component of the spin magnetic moment is reduced to $\sim H_\parallel /H_{\rm SO}$. Pair breaking occurs 
when the modified Zeeman energy $\sim (H_\parallel^2 /H_{\rm SO})\mu_B$ (known as van Vleck paramagnetism) overcomes the superconducting gap. 
For strong SOC, an Ising superconductor therefore exhibits an enhanced upper critical field $H_{{\rm cr},||}\simeq \sqrt{H_pH_{\rm SO}}$.

\section{Summary and Outlook}

Research into 2D crystals, a field born with the discovery of graphene, is  growing at a fast pace with each new atomically thin material that is isolated or synthesised. Here, we have reviewed the main features of the currently most studied 2D crystals, including metals (NbSe$_2$), semimetals (graphene), semiconductors (TMDs, phosphorene, etc.) and insulators (h-BN). Individually or combined with other layered materials to form van der Waals heterostructures, 2D materials are demonstrating new physics of fundamental interest, but also with significant potential for future applications in photonics and nanoelectronics. Some of the new possibilities discussed in this tutorial include the manipulation of spin and valley degrees of freedom for application in spintronics and valleytronics, the generation and confinement of excitons, the tuning of the optoelectronic properties of 2D materials by strain engineering, or the emergence of novel superconducting phases, such as Ising superconductivity, only possible in 2D. The intense activity in this field of research promises the discovery of new exciting phenomena, that will continue to demonstrate the unique physics of crystalline solids as they are reduced to a two-dimensional, atomically thin form.

\section*{Acknowledgements}

We thank A. N. Rudenko and J. L. Lado for kindly providing band structures of phosphorene, antimonene and MoS$_2$. We acknowledge financial support from the Spanish Ministry of Economy and Competitiveness through Grant Nos. FIS2014-58445-JIN, FIS2015-65706-P and FIS2016-80434-P (MINEICO/FEDER), the Ram\'on y Cajal programme, Grant Nos. RYC-2011-09345 and RYC-2013-14645, the ``Mar\'ia de Maeztu'' Programme for Units of Excellence in R\&D (MDM-2014-0377), the European Union's Seventh Framework Programme (FP7/2007-2013) through the ERC Advanced Grant NOVGRAPHENE (GA No. 290846) and European Commission under the Graphene Flagship, contract CNECTICT-604391.

\bibliography{rsc} 
\bibliographystyle{rsc} 

\end{document}